\documentclass[aps,pra,showpacs,twocolumn,superscriptaddress,floatfix,fleqn,byrevtex]{revtex4}
\usepackage{amsmath,amssymb}
\usepackage{graphics,color,epsfig}

\begin{document}
\title{Optical switching and inversionless amplification controlled by
state-dependent alignment of molecules}
\author{A.~K.~Popov} \altaffiliation [Corresponding author:]
{Alexander Popov}\email{apopov@uwsp.edu}
\homepage{http://www.kirensky.ru/popov} \affiliation{ Department
of Physics \& Astronomy and Department of Chemistry, University of
Wisconsin-Stevens Point, Stevens Point, WI 54481, USA}
\affiliation{Institute of Physics of the Russian Academy of
Sciences, 660036 Krasnoyarsk, Russia}
\author{V.~V.~Slabko}
\email{slabko@iph.krasn.ru} \affiliation{Institute of Physics of
the Russian Academy of Sciences, 660036 Krasnoyarsk,
Russia}\affiliation{Krasnoyarsk State Technical University, 660074
Krasnoyarsk, Russia}
\date{January 5, 2005}
\begin{abstract}
Switching anisotropic molecules from strongly-absorbing to
strongly-amplifying  through a transparent state is shown to be
possible by application of dc or ac control electric fields
without the requirement of the population inversion. It is based
on decoupling of the lower-level molecules from the resonant light
while the excited ones remain emitting due to their
state-dependent alignment. The amplification index may become
dependent only on a number of excited molecules and not on the
population of the lower state. A suitable class of molecules and
applications in optoelectronics, fiberoptics and nanophotonics are
outlined.
\end{abstract}
\maketitle
\section{Introduction}
The feasibility of manipulating transparency of materials from a
strongly-absorbing state to amplifying via absolute transparency
without a change of the populations of the coupled energy-levels
is an alluring prospect because of numerous important
applications. Amplification of light is determined by the
difference between the net absorption and stimulated emission of
radiation. Usually, this requires a larger population of the upper
energy level than the lower one. However, the amount of the
absorbed and emitted photons depends not only on the populations
of the resonant energy levels but also on the probabilities of the
induced transitions and on the distribution of the populations
over the energy-degenerated states. Thus, population inversion is
not the only option for the amount of coherently-emitted light to
prevail over the absorbed one. A variety of processes that may
lead to nonreciprocity of emission and absorption were
investigated in the early years of quantum electronics. Among them
are those based on the difference in velocity distributions in the
upper and lower states \cite{{65a},{65b}}, on nonlinear
interference effects (NIE) at quantum transitions  in the field of
the auxiliary radiation \cite{{Nie1},{Nie2}} in two-level atoms
\cite{RSob}, and the related feasibility of inversionless
amplification in three-level atoms \cite{Nrec}. The feasibility of
manipulating the transmission of light from attenuation to
amplification without population inversion (AWI) via transparency
based on NIE in three-level atoms was investigated in
\cite{{LWI1},{LWI2},{LWI3}} and illustrated with numerical
examples for transitions of Ne driven by  a He-Ne laser at the
adjacent transitions. It was experimentally proved for the
transition 2p$_4$--2s$_2$ of Ne probed by the $\lambda$ =
1.15$\mu$ He-Ne laser,  whereas  the adjacent transition
2s$_2$--2p$_1$ was driven by another $\lambda$ = 1.52$\mu$ He-Ne
laser \cite{Bet}. Possible nonreciprocity at the transitions to
the autoionizing states and related AWI were discussed in
\cite{AG}. The feasibility of AWI for short pulses in a
bichromatically excited double-lambda scheme was investigated in
\cite{KKh}. A review of later publications on AWI is given in
\cite{Corb}. An asymmetry in the absorption and emission
lineshapes of a two-level system caused by the interaction with a
thermostat is discussed in \cite{Sh}. In the present paper, a
different type of nonreciprocity in the absorption and emission
processes is explored. It is associated with the
energy-level-selective alignment of molecules \cite{PS1,PS2}. An
efficient  optical switching from the strong absorption to the
transparency and to AWI is shown to be achieved through decoupling
of the lower-level molecules from polarized resonant radiation,
while the upper-level molecules remain amplifying. A suitable
class of molecules and applications in optoelectronics,
fiberoptics, and nanophotonics are discussed. The proposed
feasibilities are illustrated with numerical models.
\section{Absorption index controlled by external electric field}
In the dipole approximation, the probability of induced
transitions between an upper level $m$ and lower one $g$,
$w_{mg}$, depends on the squared modulus of the projection of the
dipole transition matrix element on a vector of the resonant
electric field $\mathbf{E}$ which causes such a transition,
$|d_{mg}|^2\cos^2 \theta $,
\begin{equation}
w_{mg}={B}|E|^2F(\omega),\quad {B}=8\pi |d_{mg}\cos
\theta|^2/\hbar^2.\label{w}
\end{equation}
Here, ${B}$ is the Einstein coefficient, $F(\omega)$ is the
frequency dependence (the transition spectral form-factor), and
$\omega$ is frequency of radiation. For a molecule, the angle
$\theta$ is determined  by its orientation in space and by its
symmetry. Under the influence of an auxiliary control electric
field $\mathbf{E}_0$, the molecule turns towards the direction of
the minimum of the interaction energy. Hence, the degree of
alignment of a molecular medium depends on the alignment
parameter, which is given by the ratio of interaction energy with
the control field in the quantum state $j$ to the energy of
thermal motion $kT$ that renders disorientation. \emph{However,
the energy of such interaction $U_j$, and consequently the degree
of orientation, can be different for molecules in the lower and
upper energy levels.} Therefore, in such a case, \emph{the
probabilities of induced emission and absorption of polarized
light, averaged over the molecules with different orientations,
are not equal}. This enables inversionless amplification of the
polarized light through manipulating the difference in the
orientation-degree of the upper- and lower-level molecules with
the aid of the auxiliary dc or ac control fields without a change
in their energy-level populations.

In the presence of a control field, the amplification index
$\alpha>0$ for a linear-polarized probe radiation, $\mathbf{E}$,
is given by
\begin{equation}
\alpha=\sigma_0\int [n_mf_m(\Theta, \mathbf{E}_0)-n_gf_g(\Theta,
\mathbf{ E}_0)]\cos^2 \theta d\Theta.\label{a}
\end{equation}
Here, $n_m$ is the orientation-integrated population of the upper
and $n_g$ of the lower level, $d\Theta=\sin \theta d\theta d\phi$
is the element of solid angle, and $\sigma_0=8|\pi d_{mg}|^2\omega
F(\omega)/c\hbar$ is the absorption cross-section for  molecules,
whose transition dipole moment is aligned along  $\mathbf{E}$.
Excitation of molecules can be accomplished in any common way,
such as incoherent excitation or others employed, e.g., for dye
and solid-state lasers. The functions
$f_{m,g}(\Theta,\mathbf{E}_0)d\Theta$ depict the distributions of
molecules over orientations in the corresponding levels. It is
readily seen that \emph{their difference provides the feasibility
of AWI}. If the energy level lifetime excesses significantly  the
time required to set the orientation balance, they are given by
the Boltzmann distribution
\begin{equation}
f_j(\Theta, \mathbf{E}_0)=A_j\exp\{-U_j
(\Theta,\mathbf{E}_0)/kT\}.\label{f}
\end{equation}
Here, $A_j^{-1}=\int\exp\{-U_j(\Theta,\mathbf{E}_0)/kT\}d\Theta$
is a scale factor, T - temperature, k -  Boltzmann constant, and
$j = \{g, m\}$. The potential energy of a molecule coupled with
the electric field $\mathbf{E}_0$ can be presented as
\cite{Kie1,Kie2}
\begin{equation}
U_j = -\mu_i^jE_{0i}-\alpha_{ik}^jE_{0k}^*E_{0i}/2.\label{u}
\end{equation}
Here, $\mu_i^j$ is the $i^{th}$ component of the permanent dipole
moment, and $\alpha_{ik}^j$ is the component of the tensor of
electrical polarizability, both for the molecules in the energy
level $j$; and $E_{0i}$ is the $i^{th}$ component of the control
field. The first term in Eq. (\ref{u}) represents the energy of
molecules without a center of symmetry, which possess a permanent
dipole moment. The second term depicts the interaction energy with
the dipole induced by the field $\mathbf{E}_0$. \emph{This term
presents the alignment caused  by either a strong dc or ac}, e.g.,
laser field. For a review of the laser alignment, including recent
experiments, see \cite{Stap,Exp1,Exp2,Exp3,Hol} and references
therein.

Consider  axial-symmetric molecules and further assume the
directions of both the permanent $\mathbf\mu$ and  the transition
moment $\mathbf{d}_{mg}$ are aligned along the symmetry axis of
the molecule, which makes an angle $\theta_0$ with the control
field $\mathbf{E}_0$. Then the interaction energy (\ref{u}) and
distribution functions (\ref{f}) become dependent only on
$\theta_0$:
\begin{equation}
f_j(\Theta, \mathbf{E}_0)=f_j(\theta_0,
{E}_0)=A_j\exp\{p_j\cos\theta_0+ q_j\cos^2\theta_0\}.\label{f}
\end{equation}
Following \cite{Kie1,Kie2}, we introduce the alignment parameters
attributed to the permanent dipole orientation $p_j$ and to the
polarizability ellipsoid $q_j$ as
\begin{equation}
p_j=\mu^jE_0/kT,\quad
q_j=(\alpha^j_{33}-\alpha^j_{11})|E_{0}|^2/2kT. \label{pq}
\end{equation}
Here, $\alpha^j_{33}$ and $\alpha^j_{11}$ are the principal values
of the polarizability tensor along and across the symmetry axis,
accordingly, for the molecule in level $j$. They are  calculated
as \cite{Kie1,Kie2}
\begin{equation}
\alpha^j=\frac{2}{\hbar}\sum_l
\frac{\omega_{lj}|d_{lj}|^2}{\omega^2_{lj}-\omega^2_0}=\frac{e^2}{m}\sum_l
\frac{f_{lj}}{\omega^2_{lj}-\omega^2_0}, \label{b}
\end{equation}
where $d_{lj}$ is the transition dipole moment along or across the
molecule symmetry axis accordingly between the corresponding
levels, $f_{lj}$ is the corresponding oscillator strength, $m$ and
$e$ are the electron mass and charge, $\omega_{lj}$ is transition
frequency, and $\omega_0$ is the frequency of $\mathbf{E}_0$.
\section{Molecules with large ground-state
permanent dipole moment} Consider axial-symmetric molecules whose
alignment is primarily determined by the permanent dipole moment.
If such a moment in the lower state is larger than in the upper
one ($\mu_g>\mu_m$), the alignment degree in the lower state is
larger than for molecules in the upper state. Consequently, the
orthogonal orientation of the probe $\mathbf{E}$ and control
$\mathbf{E}_0$ fields is advantageous for suppression of the
absorption and for enhancement of the emission. The
orientation-averaged absorption/amplification index in the case of
orthogonal orientation of the probe and the control fields is
calculated as
\begin{align}
\frac{\alpha_{\perp}}{\sigma_0n_g}&=\int^{2\pi}_0 d\varphi
cos^2\varphi \int^{\pi}_0d\theta_0
sin^3\theta_0\nonumber\\
&\times\left[\frac{n_m}{n_g}A_m\exp\{p_m
\cos\theta_0\}-A_g\exp\{p_g
\cos\theta_0\}\right]\nonumber\\
&=\frac{n_m}{n_g}\frac{L(p_m)}{p_m}-\frac{L(p_g)}{p_g},\label{o}
\end{align}
where  $L(p_j)$ is the Langevin function
\cite{Kie1,Kie2,Jahn1,Jahn2}
\begin{equation}\begin{split}
L(p_j)&=(1-\frac{2}{p_j})^{-1}\nonumber\\
&\times \int^{2\pi}_0 d\varphi\int^{\pi}_0d\theta_0 A_j\exp\{p_j
\cos\theta_0\}\cos^2\theta_0\sin\theta_0\nonumber\\
&=\frac{\int^{\pi}_0d\theta_0 \exp\{p_j
\cos\theta_0\}\cos\theta_0\sin\theta_0}{\int^{\pi}_0d\theta_0
\exp\{p_j \cos\theta_0\}\sin\theta_0}=\coth
p_j-\frac{1}{p_j}.\end{split}\label{ft}
\end{equation}

\begin{figure}[!h]
\includegraphics[width=.5\textwidth]{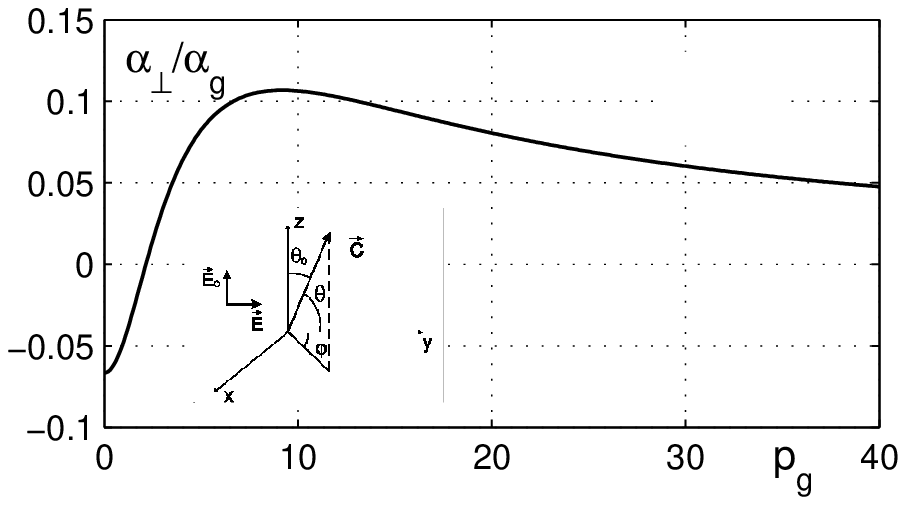}
\caption{\label{f1} Scaled amplification index
$\alpha_{\perp}/\alpha_g$ ($\alpha_g = n_g \sigma_0$) vs alignment
parameter~$p_g = \mu_gE_0/kT$. $n_m = 0.8 n_g$.  $\mu_g = 4
\mu_m$. The orientation of the molecule symmetry axis $\mathbf C$
is depicted in the inset.}
\end{figure}

\begin{figure}[!h]
\includegraphics[width=.48\textwidth]{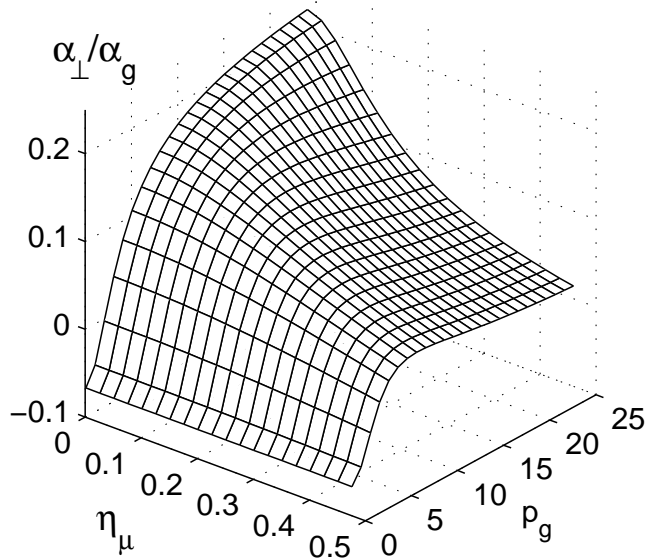}
\caption{\label{f2}Scaled amplification index vs alignment
parameter $p_g = \mu_gE_0/kT$ and  ratio of the excited- and
ground-state permanent dipole moments $\eta_{\mu} = \mu_m/\mu_g$
for the case of $\mu_g > \mu_m$ and orthogonal orientation of the
control and probe fields. $n_m = 0.8 n_g$.}
\end{figure}

\begin{figure}[!h]
\includegraphics[width=.45\textwidth]{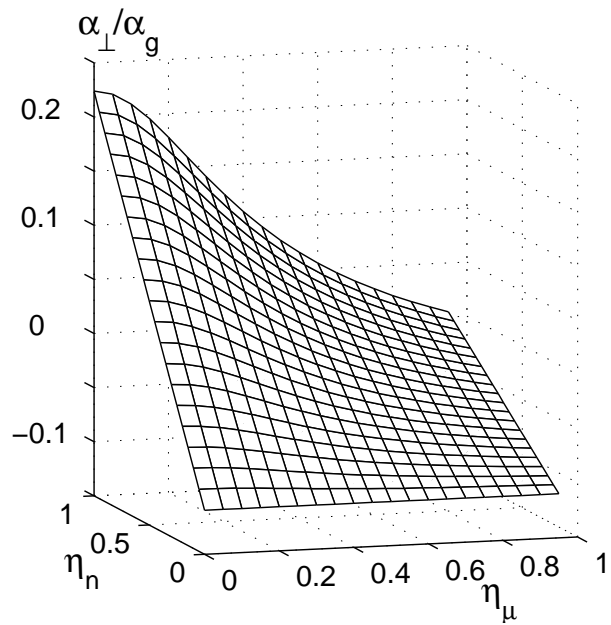}
\caption{\label{f3}Scaled amplification index vs ratio of the
excited and ground-state permanent dipole moments
$\eta_{\mu}=\mu_m/\mu_g$ and  population ratio $\eta_{n}=n_m/n_g$
for the case of $\mu_g>\mu_m$ and orthogonal orientation of the
control and probe fields. $p_g$ = 8}
\end{figure}

Figures \ref{f1} -- \ref{f3} depict features of the optical
switching from absorption to inversionless amplification in this
case.
Figure \ref{f1} illustrates the case where  the upper-state
permanent dipole moment, $\mu_m$, is four times smaller than the
one in the lower sate, $\mu_g$, $\mathbf{E}\perp\mathbf{E}_0$, and
the population of the upper energy level is 20\% less than that of
the lower one. It shows that the absorption decreases with an
increase of the control field ${E}_0$ and with a decrease of
temperature. The sample becomes transparent in the range of the
orientation parameter $p_g$ between  1 and 2. Then the probe
radiation is amplified at any strength of $E_0$ above this
threshold value (see also Fig.~\ref{f2}). The threshold value
depends on the population ratio, $n_m/n_g$ (see Fig.~\ref{f3}).
The amplification index reaches its maximum for the interval of
$p_g$ between 5 and 10. In the given specific case, the maximum
value of the amplification factor makes up
$\alpha_{max}\approx0.1\sigma_0n_g=0.1\sigma_0n/1.8$, where $n$ is
the total molecular number density. Because the absorption index
at the control field and the excitation of the upper level turned
off is $\alpha(E_0=0)=-n\sigma_0/3$, this corresponds to a pretty
strong AWI. Maximum values of AWI grow with a decrease of the
ratio $\mu_m/\mu_g$ (Fig.~\ref{f2}) and with an increase of the
amount of excited molecules, $\eta_{n}$ (see Fig.~\ref{f3}).

At $\mu_m=0$ and $ p_g\rightarrow\infty$ equation (\ref{o})
reduces to
\begin{equation}
\alpha_{\perp}=n_m\sigma_0/3,\quad
(\alpha_{\perp}/\sigma_0n_g=\eta_n/3).\label{ad}
\end{equation}
In this case, \emph{amplification is determined only by the
upper-state molecules and does not depend on the population of the
lower state}. This is because the lower-state molecules are
aligned orthogonal to the the probe field and, therefore, are not
coupled with this field. On the contrary,  the excited molecules
are decoupled from the control field $E_0$. Their orientation
remains isotropic and, hence, the averaged projection of the
transition dipole moment $\mathbf{d}_{mg}$ on the probe field
$\mathbf{E}$ is not equal to zero.
\section{Molecules with large excited-state
permanent dipole moment} In the alternative case, $\mu_m>\mu_g$,
the polarization $\mathbf{E}$ along $\mathbf{E}_0$ appears
advantageous because it ensures  stronger coupling of the probe
field with the upper-state molecules than with the ground-state
ones. The orientation-averaged absorption/amplification index is
calculated accordingly as
\begin{align}
\frac{\alpha_{\upuparrows}}{\sigma_0n_g}&= \int^{2\pi}_0 d\varphi
\int^{\pi}_0d\theta_0 cos^2\theta_0
sin\theta_0\nonumber\\
&\times\left[\frac{n_m}{n_g}A_m\exp\{p_m
\cos\theta_0\}-A_g\exp\{p_g
\cos\theta_0\}\right]\nonumber\\
&=\frac{n_m}{n_g}\left[1-2\frac{L(p_m)}{p_m}\right]-\left[1-2\frac{L(p_g)}{
p_g}\right].\label{p}
\end{align}

\begin{figure}[!h]
\includegraphics[width=.5\textwidth]{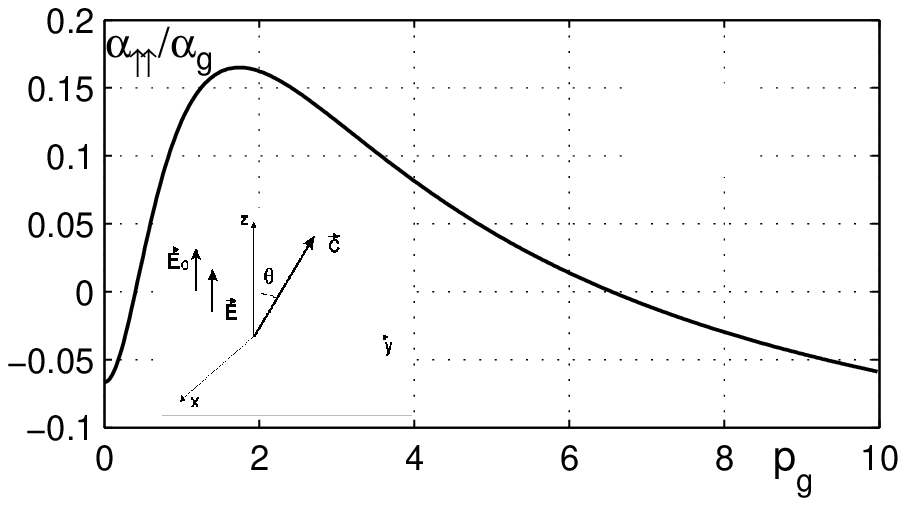}
\caption{\label{f4} Scaled amplification index
$\alpha_{\upuparrows}/\alpha_g$ ($\alpha_g = n_g\sigma_0$) vs
alignment parameter~$p_g$ for $n_m = 0.8 n_g$. $ \mu_m = 4 \mu_g$.
The orientation of the molecule symmetry axis $\mathbf C$ is
depicted in the inset.}
\end{figure}
\begin{figure}[!h]
\includegraphics[width=.48\textwidth]{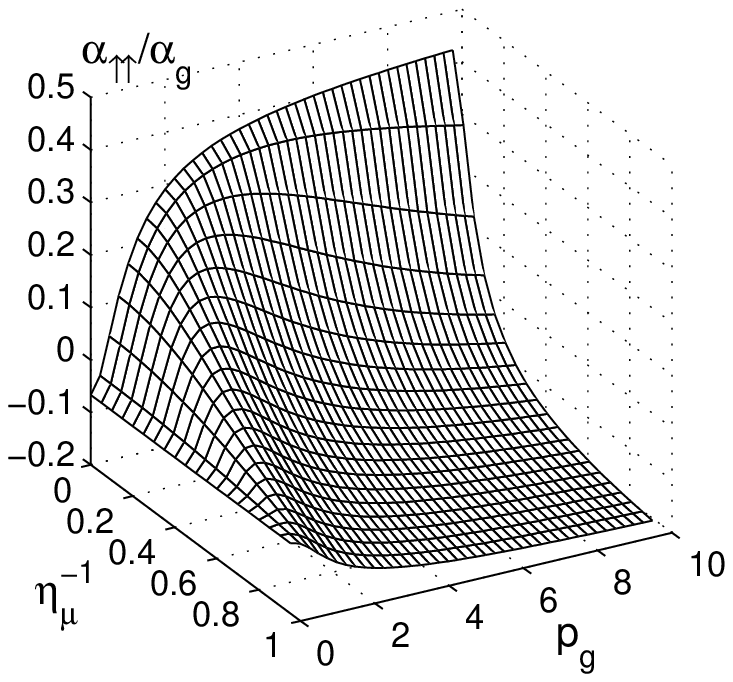}
\caption{\label{f5}Scaled amplification index vs alignment
parameter $p_g = \mu_gE_0/kT$ and ratio of the excited- and
ground-state permanent dipole moments $\eta_{\mu}^{-1} =
\mu_g/\mu_m$ for the case of $\mu_m>\mu_g$ and parallel
orientation of the control and probe fields.  $n_m = 0.8 n_g$. }
\end{figure}
\begin{figure}[!h]
\includegraphics[width=.45\textwidth]{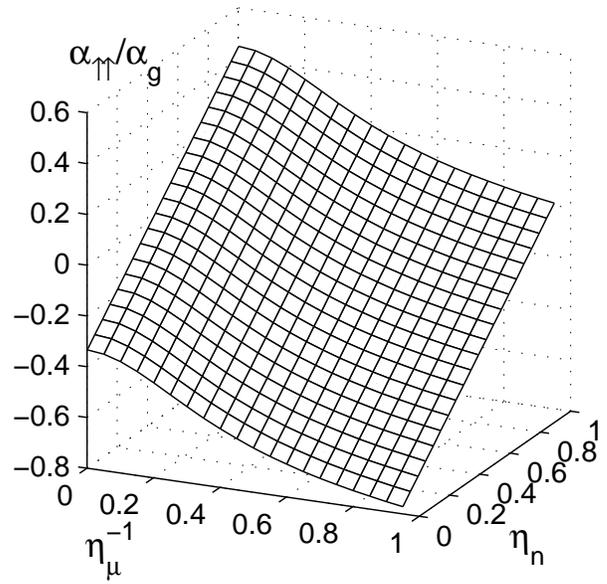}
\caption{\label{f6}Scaled amplification index vs ratio of the
excited- and ground-state permanent dipole moments
$\eta_{\mu}^{-1}=\mu_g/\mu_m$, and population ratio
$\eta_{n}=n_m/n_g$ for the case of $\mu_m>\mu_g$ and parallel
orientation of the control and probe fields. $p_g$ = 2.}
\end{figure}
Figures \ref{f4} -- \ref{f6} depict the switching features  in
this case, which include inversionless amplification. They display
a different dependence of the amplification index on the alignment
parameter and on the upper-state populations than in the previous
case. Figure \ref{f4} depicts the case when the permanent electric
dipole moment in the upper state is four times larger than that in
the lower state. All other parameters remain the same as in
Fig.~\ref{f1}.  Besides switching from absorption to amplification
within relatively narrow interval of the small alignment
parameters $p_g\approx0.05$, Fig. \ref{f4} displays a second
interval of more flat switching behavior in the vicinity of
$p_g\approx6.5$. Figures \ref{f5} and \ref{f6} show that with
$\mu_g\rightarrow0$, the amplification may become comparable with
the absorption corresponding to the control field and excitation
of the upper level turned off. The figures also show that the
transparency or one and the same magnitudes of the amplification
index can be achieved for different sets of population and
alignment parameters.
\section{Molecules without permanent dipole
moment} Figures \ref{f1} -- \ref{f6} depict optical switching with
a dc control electrical field for molecules that possess permanent
electric dipole moment. Similar conclusions are applied to optical
switching and AWI induced by an ac field $E_0$ due to the
difference of the induced dipole moments (polarizabilities) in the
upper and lower states for axially-symmetric molecules without
permanent electric dipole moment. In such case, equations
(\ref{o}) and (\ref{p}) take the form
\begin{align}
\frac{\alpha_{\perp}}{\sigma_0n_g}&=\int^{2\pi}_0 d\varphi
cos^2\varphi \int^{\pi}_0d\theta_0
sin^3\theta_0\nonumber\\
&\times\left[\frac{n_m}{n_g}A_m\exp\{q_m
\cos^2\theta_0\}-A_g\exp\{q_g
\cos^2\theta_0\}\right]\nonumber\\
&=\frac{n_m}{n_g}[1-{L_2(q_m)}]-[1-{L_2(q_g)}], \label{oq}\\
\frac{\alpha_{\upuparrows}}{\sigma_0n_g}&= \int^{2\pi}_0 d\varphi
\int^{\pi}_0d\theta_0 cos^2\theta_0
sin\theta_0\nonumber\\
&\times\left[\frac{n_m}{n_g}A_m\exp\{q_m
\cos^2\theta_0\}-A_g\exp\{q_g
\cos^2\theta_0\}\right]\nonumber\\
&=\frac{n_m}{n_g}{L_2(q_m)}-{L_2(q_g)}.\label{pq}
\end{align}
Here, $L_2(q)$ is the generalized Langevin function
\cite{Kie1,Kie2,Jahn1,Jahn2}
\begin{align}
L_2(q)=&\frac{\int^{\pi}_0d\theta_0 cos^2\theta_0
sin\theta_0\exp\{qcos^2\theta_0\}}{\int^{\pi}_0d\theta_0
 sin\theta_0\exp\{q\cos^2\theta_0\}}\nonumber\\
 =&\left[2\sqrt{q}e^{-q}\int^{\sqrt{q}}_0dt\exp\{t^2\}\right]^{-1}-\frac{1}{2q}.\label{gl}
\end{align}
Parallel orientation of polarizations of the control and probe
fields is advantageous for molecules with larger parameter q in
the upper state ($q_{g}<q_m$), whereas the orthogonal orientation
is advantageous in the opposite case, $q_{m}<q_g$.

The important feature of ac alignment is that the sign of the
parameter q depends not only on the structure of the molecules but
on the ratio of frequency $\omega_0$ of the control field and that
of the strongest transitions contributing to the alignment,
$\omega_{ij}$. For molecules stretched along the symmetry axis,
$\alpha_{33}>\alpha_{11}$, and $q>0$ at $\omega_{0}<\omega_{ij}$.
Alternatively, for disc-type molecules, $\alpha_{33}<\alpha_{11}$,
and $q<0$ at $\omega_{0}<\omega_{ij}$. Consequently, in the latter
case, the control field decreases the orientation-averaged
projection of the induced moment on the vector of the control
field. Therefore, the parallel orientation
$\mathbf{E}_0\upuparrows\mathbf{E}$ of the probe and control
fields is optimal for the cases of $0<q_g<q_m$ and $q_g<q_m<0$,
but in the first case the dependence $\alpha(|E_0|^2)$ resembles
Fig. \ref{f4}, whereas in the second case -- Fig.~\ref{f1}. In the
opposite case, $q_{g}>q_m>0$ and $0>q_{g}>q_m$, the orthogonal
orientation of the fields $\mathbf{E}_0\perp\mathbf{E}$ is
advantageous, and the dependence $\alpha(|E_0|^2)$ is
qualitatively analogues to Fig. \ref{f1}, whereas in the second
case -- to Fig.~\ref{f4}. The effect is most strong when the signs
of q are different in the upper and lower states. This may happen
if the signs of the detuning of $\omega_0$ relative to the
transitions dominantly contributing to the alignment are different
for molecules in the upper and lower states. Then the control
field aligns molecules along the direction of its polarization in
one state, whereas there is a decrease of the number of such
molecules in another coupled state.
\section{Suitable molecules and applications}
A  variety of dye-type molecules possess state-dependent permanent
dipole moments and state-dependent polarizability tensors. For
example, the molecule 3-6-diacetyl-amino-phtalimide does not
possess a permanent dipole moment in the excited state
($\mu_m=0$), whereas the ground state has $\mu_g\approx5.5$~D
\cite{Ter}. The important requirement for realization of the
proposed technique is that the lifetime in the excited state
$\tau$ must exceed the time $\tau_0$ necessary to set the
orientation equilibrium. The alignment time varies depending on
the molecular size and on the viscosity of the solvent. For large
organic molecules in solvents, the characteristic alignment time
is $\tau_0\simeq 10^{-10} - 10^{-12}$~s, for protein
macromolecules it is $\tau_0\simeq 10^{-6} - 10^{-8}$ s, and
$\tau_0\simeq 10^{-2} - 10^{-4}$~s for big biomacromolecules.
Hence, for the applications under consideration, the molecular
mass must not significantly exceed $10^3$ atomic units, and their
length should be of several nanometers. For such compounds, a
permanent electric dipole moment is usually on the order of
$1-10$~D \cite{Kie1,Kie2,Ter,Bak,Ver}. In the
electrically-resistant solvents, an alignment parameter of several
units can be achieved for such molecules with a dc field $E_0$
below the breakdown threshold \cite{Kik}. The breakdown threshold
can be increased by the application of pulsed dc fields $E_0$ and
by a decrease of the temperature of the solvents, which is
possible by using cryogenic liquids or cooled gas. For example, at
$\mu_g=10$~D, T = 70 K, the parameter $p=1$ is achievable at
$E_0\approx 28$~kV/mm. Such a strength of the control field is
within the typical range used in laser Stark spectroscopy
\cite{Dem} and can be easily realized, e.g., in hollow fibers. A
greater alignment is achievable for molecules having dipole moment
on the order of $10^2 - 10^3$~D. However, their reorientation time
$\tau_0$ may be comparable or longer than the lifetime in the
exited state, which reduces the effect. Thus, optical switching
and AWI may appear as easier to achieve through orientation of
molecules by an ac optical field. The polarizability anisotropy
$\alpha_{33} - \alpha_{11}$ is $\sim 10^{-23}$ cm$^3$  even for
large resonance detunings
$|\omega_{ij}-\omega_{0}|\sim\omega_{ij}$. At a normal temperature
T = 300 K,  the optical alignment parameter reaches the magnitude
$|q|\sim1$ with the control radiation of $P\sim$ 10~-~100~MW
focused below the breakdown threshold. The magnitude of $q$ can be
considerably increased and the strength of the required control
field decreased by setting the frequency $\omega_0$ in the
vicinity of the resonant frequencies of the adjacent optical
transitions. Alignment of about 70\% of molecules
($<\cos^2\theta>\sim 0.75$) has been recently achieved in
experiments with short laser pulses \cite{DAMOP}, and various
important applications in optical physics are foreseen
\cite{Stap}. A possible alignment of linear molecules in hollow
fibers is shown in \cite{Hol}.

Besides hollow fibers, the proposed technique can be implemented
in nanophotonics and near-field optics, especially associated with
plasmonic metamaterials. Such nanostructures as metal islands on a
dielectric surface or colloidal metal fractal aggregates provide a
presence of nanometer-scale spatial regions of very high
concentration of local electric and electromagnetic fields, i.e.,
``hot'' spots. Accordingly, this leads to many-order enhancements
for a variety of optical processes, such as photoluminescence,
nonlinear absorption and refraction, Raman scattering and
four-wave mixing. This facilitates new possibilities for optical
microanalysis, spectroscopy of single molecules, and lasing and
nonlinear optical processes down to a single molecule. The
features associated with the inhomogeneity of the near-fields and
with the alignment of molecules adsorbed onto metal nanostructures
are now opened for investigation, where the experiments carried
out to date have provided encouraging results. Giant enhancement
of optical  effects including lasing at an ultra-low threshold and
enhancement of the excitation efficiency up to $10^{11}$ have been
observed in rhodamine 6G molecules added to silver colloidal
nanoaggregates which were placed inside the quartz microcavity
\cite{Vsh1,Vsh2,Vsh3,Vsh4}.
\section{Conclusions}
In summary,  we propose to facilitate the optical switching from
strong absorption to transparency, and further to amplification
and lasing without the population inversion by the anisotropic
molecules through a decoupling of a polarized light from the
absorbing molecules in the lower energy level while the
upper-state molecules stay emitting. Such switching becomes
feasible with the aid of a control dc electric or an ac optical
field that aligns anisotropic molecules in upper- and lower-energy
levels in a different way. The example of a suitable molecule,
similar to dye molecules, is given, and the favorable conditions
are discussed and numerically illustrated. The achievable
magnitudes of amplification without population inversion may reach
values that are determined only by the excited-states molecules
and do not depend on the population difference. The outlined
features promise novel applications of the alignment, such as in
optoelectronics, fiberoptics  and nanophotonics. Among them are
advances in the design of optical micro- and nanoswitches and
nanolasers operating on a small number of (or even on individual)
molecules adsorbed on metal nanostructures that are placed inside
a micro- or nanocavity or inside a hollow fiber. This offers novel
applications for control over the combined evanescent and
surface-enhanced radiative processes with high-Q
morphology-dependent resonances by the auxiliary dc or optical
electric fields.
\begin{acknowledgments}
We thank S.A. Myslivets for help with the numerical analysis. This
work was supported in part by DARPA under grant MDA972-03-1-0020.
\end{acknowledgments}

\end{document}